%% file: main.tex
\documentclass[a4paper,11pt]{article}
\usepackage{pos}
\usepackage{soul}

\title{Searching for VHE gamma-ray emission associated with IceCube neutrino alerts using FACT, H.E.S.S., MAGIC, and VERITAS}
 \ShortTitle{IceCube-IACT neutrino ToOs}

\author*[a]{Konstancja Satalecka}
\author[b]{Elisa Bernardini} 
\author[c]{Daniela Dorner}
\author[d]{Ga\v{s}per Kukec Mezek}
\author[e]{Weidong Jin}

\affiliation[a]{DESY, Platannenallee 6, 15387 Zeuthen, Germany}
\affiliation[b]{University of Padova,  Physics \& Astronomy Dep. "G. Galilei",
F.  Marzolo, 8, 35121, Padova, Italy}
\affiliation[c]{Universit\"at W\"urzburg, D-97074 W\"urzburg, Germany}
\affiliation[d]{Linnaeus University, Department of Physics and Electrical Engineering, 35195 V\"axj\"o, Sweden}
\affiliation[e]{University of Alabama, Department of Physics and Astronomy, Tuscaloosa, AL, USA\newline}

\forColl{MAGIC, IceCube, FACT, H.E.S.S., VERITAS} 
\emailAdd{konstancja.satalecka@desy.de}
\emailAdd{elisa.bernardini@unipd.it}
\emailAdd{dorner@astro.uni-wuerzburg.de}
\emailAdd{gasper.kukecmezek@lnu.se}
\emailAdd{wjin4@crimson.ua.edu}

\abstract{The realtime follow-up of neutrino events is a promising approach to search for astrophysical neutrino sources. It has so far provided compelling evidence for a neutrino point source: the flaring gamma-ray blazar TXS 0506+056 observed in coincidence with the high-energy neutrino IceCube-170922A detected by IceCube. The detection of very-high-energy gamma rays (VHE, $\mathrm{E} > 100\,\mathrm{GeV}$) from this source helped establish the coincidence and constrained the modeling of the blazar emission at the time of the IceCube event.
The four major imaging atmospheric Cherenkov telescope arrays (IACTs) - FACT, H.E.S.S., MAGIC, and VERITAS - operate an active follow-up program of target-of-opportunity observations of neutrino alerts sent by IceCube. This program has two main components. One are the observations of known gamma-ray sources around which a cluster of candidate neutrino events has been identified by IceCube (Gamma-ray Follow-Up, GFU). Second one is the follow-up of single high-energy neutrino candidate events of potential astrophysical origin such as IceCube-170922A. GFU has been recently upgraded by IceCube in collaboration with the IACT groups. We present here recent results from the IACT follow-up programs of IceCube neutrino alerts and a description of the upgraded IceCube GFU system.}

\FullConference{37$^{\rm{th}}$ International Cosmic Ray Conference (ICRC 2021)\\
		July 12th -- 23rd, 2021\\
		Online -- Berlin, Germany}

\begin{document}
\maketitle

\section{Introduction}

In 2013, the IceCube Neutrino Observatory published observational evidence for the existence of extraterrestrial high-energy neutrinos \cite{IceCube:2013low}, which is now well established \cite{2020arXiv201103545A}. It indicates that astrophysical sources of high-energy neutrinos exist, but their identity has not yet been revealed. In general, the direct identification of neutrino sources is challenging due to low signal statistics ($\sim$10/year) and the relatively large angular uncertainties of neutrino events ($\sim$few to tens of degrees). The support of electromagnetic (EM) observations is therefore crucial for this task. 
Very-high-energy (VHE, E $>$ 100\,GeV) gamma rays are a natural energy range to search for EM neutrino counterparts, as they can be produced together in cosmic sources through hadronic interactions of cosmic rays (CRs). The two standard production channels are hadronic or photo-hadronic interactions, where the resulting charged and neutral pions decay into neutrinos and gamma rays, respectively. The resulting gamma-ray spectrum at the source should have a shape and normalization very similar to the neutrino spectrum. Moreover, the highest photon energy produced is directly linked to the highest energy of the parent particle population (here protons). Thus gamma rays give us valuable insight into the CR acceleration mechanisms inside the observed source. Neutrinos only interact weakly, therefore they can travel unimpeded over large distances and dense environments. By contrast, gamma rays can be absorbed or down-scattered while passing  through their source region and during propagation across extragalactic space due to the effect of the extragalactic background light (EBL, \cite{2013APh....43..112D}). This difference introduces a strong complementarity to joint neutrino and gamma-ray observations: detection or non-detection of a VHE gamma-ray signal consistent with that of high-energy neutrinos provides diagnostic information about the distances and/or local environments of the neutrino sources.

For these reasons, all of the major currently operating Imaging Atmospheric Cherenkov Telescopes (IACTs): FACT \cite{Biland:2014fqa}, H.E.S.S.\ \cite{2006A&A...457..899A}, MAGIC \cite{2012APh....35..435A} and VERITAS \cite{Holder:2006gi} introduced observational programs aiming at neutrino event follow-up and identification of their gamma-ray counterparts. Observations of high-energy neutrino event directions, under the hypothesis of steady source emission, have been presented in \cite{2017ICRC...35..618S}. Here, we discuss follow-up of realtime IceCube neutrino alerts. This approach turned out to be successful and enabled to establish the first compelling evidence for a neutrino emitting source. In September 2017, a coincidence between the high-energy neutrino event IceCube-170922A and a gamma-ray flaring blazar TXS\,0506+056 was observed at a $\sim$3$\sigma$ level \cite{IceCube:2018dnn}. In these proceedings, we give a brief overview of the current neutrino follow-up programs lead by IACTs and their results starting from October 2017, i.e., after the IceCube-170922A and TXS\,0506+056 detection.

\section{IceCube neutrino alert channels and IACTs' follow-up strategies}

In \cite{2017APh....92...30A}, the IceCube collaboration presents a summary of its current realtime alert emission programs. From the publicly available alert channels, the high-energy ($>$60 TeV) single events of likely astrophysical origin and with well-reconstructed directions are the most interesting ones for IACTs. These events are broadcast by IceCube in realtime since 2016, with a typical latency of $\sim$30\,s. In 2019, the event selection was updated \citep{2019ICRC...36.1021B}. Events with at least 50\% probability of being astrophysical (so-called \textit{signalness}\footnote{Probability that this is an astrophysical signal relative to backgrounds, assuming the best-fit diffuse muon neutrino astrophysical power-law flux E$^{-2.19}$}) are flagged and distributed as \textit{Gold} alerts. Those with 30\% signalness are categorized as \textit{Bronze} alerts. These alerts have a localization uncertainty of $\sim1^\circ$, matching a typical IACT field of view of 3.5-5$^{\circ}$. The main goal of the IACT follow-up programs is to identify a VHE counterpart to the neutrino event. It can be, for example, an active galactic nucleus (AGN, like TXS\,0506+056) or a transient source. Without an \emph{a priori} identification of promising source candidates (e.g.,\ using the $Fermi$-LAT and IACT catalogs), the searches typically cover the whole region o finterest (ROI) defined by the neutrino localisation uncertainty. 

\begin{figure}[th]
\begin{center}
\includegraphics[width= 0.9\textwidth]{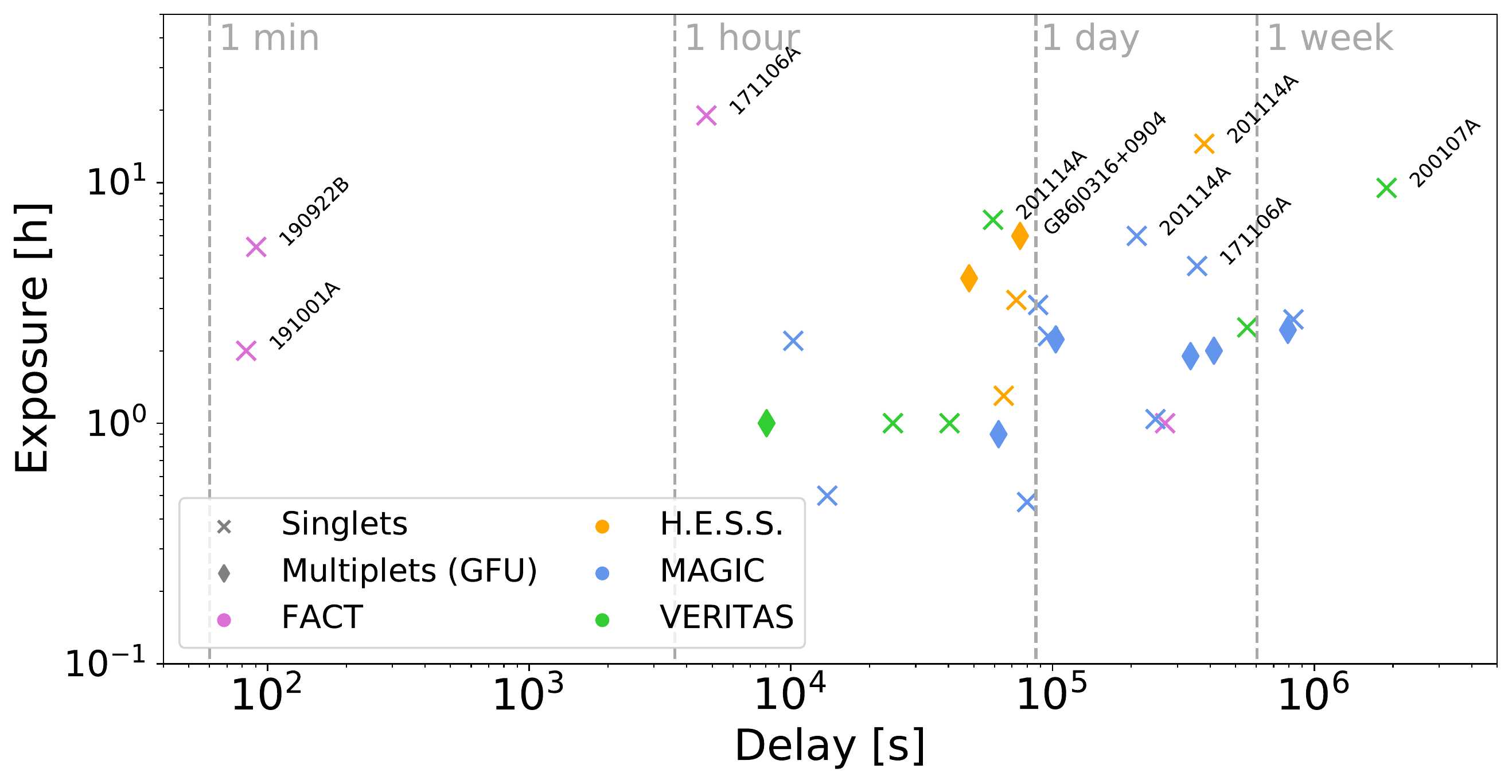}
\caption{Delay vs exposure times for IACT follow-up of neutrino alerts from October 2017 until March 2021. The delay is calculated from the neutrino event arrival time (single events) or flare threshold crossing time (multiplets) up to the start of the IACT observation. Highlighted are observations performed with a start delay less than 100\,s or with a total exposure longer than 4\,h. Marker color represents the IACT observing while the marker type represents the alert type.}
\label{fig:delay}
\end{center}
\end{figure}

If the external conditions allow it, i.e.,\ the alert arrives during a dark night, the source is visible and the weather is good, an automatic re-pointing procedure is carried out by the telescopes and the alert can be observed immediately. Otherwise, the observations typically take  place within the next few days, once these conditions are fulfilled. All telescopes introduced automatic re-pointing to the GOLD alerts (H.E.S.S. since 2016, FACT, MAGIC and VERITAS since 2019). VERITAS and FACT re-point automatically also to BRONZE alerts. Thanks to automatic re-pointing, delays between the trigger by IceCube and the follow-up observation are minimized (e.g.\ only 83~seconds for 191001A by FACT, see Fig.~\ref{fig:delay}). The first exposure ranges from 30 minutes to a few hours. Depending on the results of the first observation and available multi-wavelength information about potential EM counterparts, more data can be taken in the following nights. 

A complementary approach is used in the Gamma-ray Follow-Up (GFU) program. GFU, one of the longest operating (since 2012) neutrino follow-up programs, is dedicated for IACTs. The goal is to alert IACTs to neutrino multiplets ({\it flares}) above a pre-defined significance. The {\it flare} duration is not constrained a priori, and can range from seconds or less to 180 days. The MAGIC and VERITAS results from the first stage of the GFU program (up to 2016) are presented in \cite{2016JInst..1111009I}. Afterwards, the GFU event selection and reconstruction has been updated, with a dedicated high-energy single track stream feeding the largest fraction of the {\it Gold} and {\it Bronze} events mentioned above. Moreover,
new sources were added to the list of objects monitored for event clusters and the program was extended to the Southern hemisphere to allow follow-up observations with the H.E.S.S.\ array. Finally, an unbiased search for clusters from anywhere in the sky was developed and commissioned.

The list of objects monitored for neutrino clusters are selected as potential neutrino emitters out of the 3FGL \cite{2015ApJS..218...23A} or 3FHL \cite{2017ApJS..232...18A} catalogs based upon the following criteria:
\begin{itemize}
    \item Extragalactic source with known redshift and z $\leq$ 1.0
    \item 3FGL: variability index $>$ 77.2; 3FHL: variability based on Bayesian blocks $>$ 1
    \item Culmination at the IACT site within a chosen zenith angle limit (usually $<$45$^{\circ}$)
    \item Assuming that the source can produce a gamma-ray flare with a 10-fold increase over the average $Fermi$-LAT flux, the extrapolated flux above 100\,GeV has to exceed the IACT 5$\sigma$ sensitivity for observation times between 2.5\,h to 5\,h.
\end{itemize}
Additionally, all extragalactic sources detected by IACTs, the Galaxy Center, and the Crab Nebula have been added to the source list. The final list is comprised of  120 to 180 sources for each participating IACT. Note that the blazar TXS\,0506+056 was not in the catalog of monitored sources for this program, since prior to IceCube-170922A, its redshift was unknown. Otherwise, the GFU program would have triggered follow-up observations already in 2014.  

GFU alerts are emitted by IceCube when the neutrino {\it flare} passes a pre-defined significance threshold ($\sim$3.0 - 3.5$\sigma$ for known gamma-ray sources, depending on the choice of each IACT, and 4.2$\sigma$ for all-sky alerts) . Access to the GFU alert stream is subject to a dedicated Memorandum of Understanding (MoU) between IceCube and the IACT collaborations (currently H.E.S.S., MAGIC and VERITAS). The distribution of the alerts is handled slightly differently for each follow-up instrument. MAGIC and VERITAS receive the information via an automated e-mail message. H.E.S.S.\ is informed via a dedicated VOEvent based alert stream allowing for fully automated follow-up observations. The final decision to perform follow-up observations is taken by each IACT independently and relies typically on a combination of several factors like the intrinsic parameters of the neutrino {\it flares} (e.g.,\ false alarm rate (FAR), duration of the flare, etc.), the available visibility window, weather condition, etc. The aim of these observations is different from the ones outlined above for single-event alerts in that the associated source of the alert is known and already identified as a GeV and/or TeV emitter. The IACT observations are therefore tailored to determine the changes to the state of the source (e.g.,\ quiescence vs. flaring or spectral changes).


An overview of neutrino follow-ups is shown in Figure \ref{fig:delay}, which displays the delay and IACT exposure for all single-event alerts observed by IACTs since October 2017 (i.e.\ after IceCube-170922A) and GFU alerts from 2019-2020. The delay is calculated from the neutrino event arrival time (single events) or {\it flares} threshold crossing time (multiplets) up to the start of the IACT observation.

\section{Results}

\begin{figure}[th]
\begin{center}
\includegraphics[width= 0.9\textwidth]{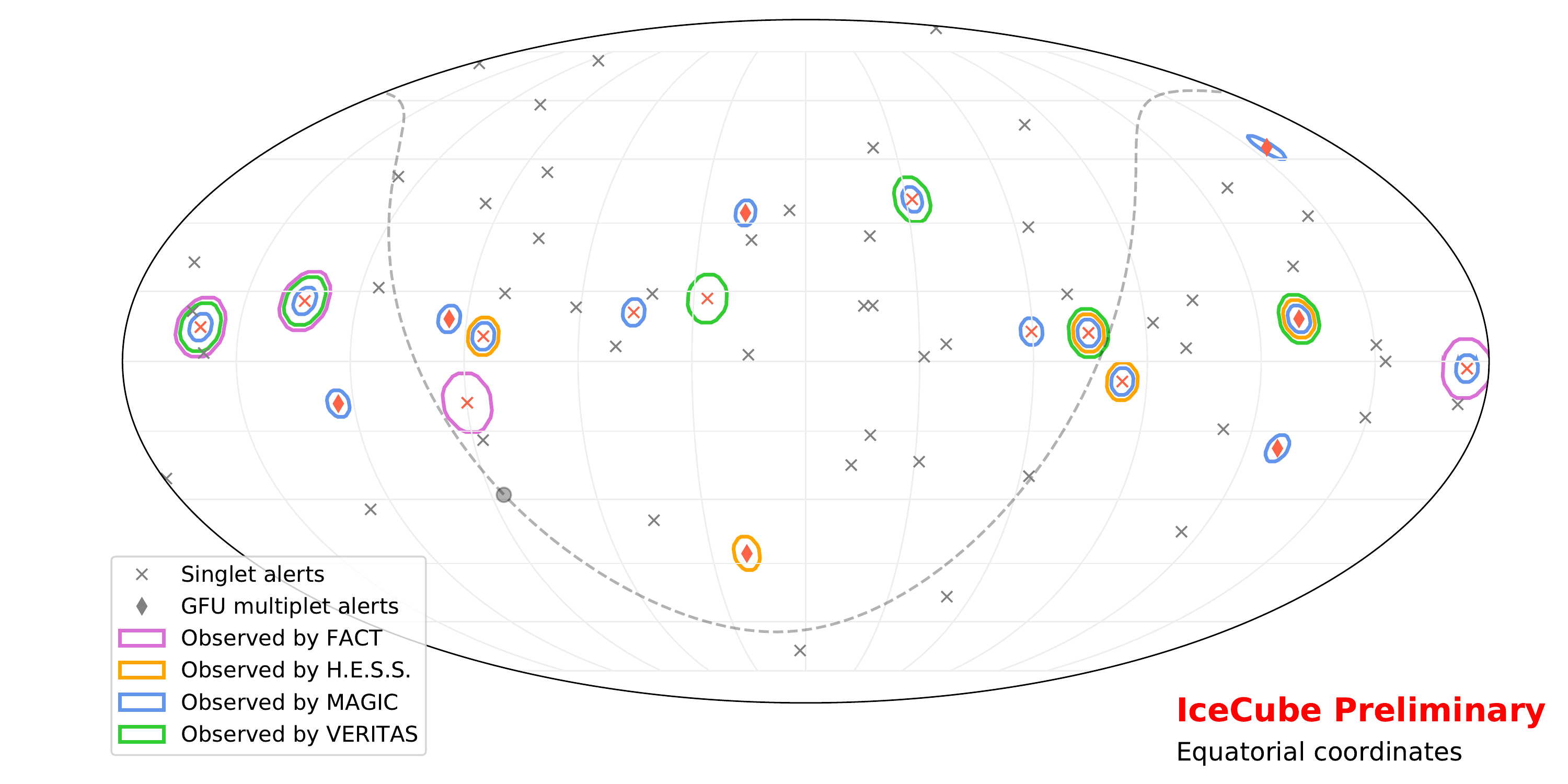}
\caption{Sky map in equatorial coordinates showing IceCube alert positions observed by IACTs between October 2017 and March 2021.}
\label{fig:alert_skymap}
\end{center}
\end{figure}

A skymap of the overview of the direction of the alerts sent by IceCube as single neutrino events and GFU neutrino {\it flares} are shown in Fig.~\ref{fig:alert_skymap}. It highlights the follow-up observations of the IACTs. 

From October 2017 until December 2020, IceCube sent 62 single event public alerts\footnote{Full list available at: \url{https://gcn.gsfc.nasa.gov/amon_icecube_gold_bronze_events.html} and following links. IceCube-200107A has been announced through a GCN Circular, several hours after its detection (see GCN \# 26655).} 
out of which 11 were observed by at least one IACT. A summary of the follow-up observations obtained by the different IACTs is given in Table~\ref{tab_tracks}. In total, each collaboration spent $\sim$20\,h of its observation time on public IceCube alerts follow-up, although with different approaches. FACT, H.E.S.S.\ and VERITAS observed 3-5 alerts each but concentrated longer exposures on a few of them (e.g.,\ FACT observed IceCube-171106A for 19 h). On the other hand, MAGIC performed the highest number of follow-ups (nine) but with a shorter average exposure. 

In the years 2019-2020, IceCube sent 27 GFU alerts from 17 sources, 7 of those sources were observed by the IACTs (cf. Table~\ref{tab_GFU}). One of them is marked as an "all-sky alert". This alert came from a channel that uses the same algorithm for flare search as GFU, but it is not restricted to the source list and therefore has a much larger number of associated statistical trials, requiring a higher pre-trial significance to issue an alert. 
The nearest source (0.35$^{\circ}$ away) to the all-sky neutrino \textit{flare} localization was PMN\,J0325-1843. 
Detailed analyses of the obtained data sets are in progress and will be the subject of a forthcoming publication. 
No significant VHE gamma-ray emission has been found in the ROIs defined by the localisation uncertainty of the single neutrino events. For the GFU alerts, no changes in source flux levels and spectra have been detected with respect to previous observations.  

 In the case of IceCube-201114A, a dedicated multi-wavelength follow-up campaign (including H.E.S.S., MAGIC and VERITAS) was organized for its potential counterpart 4FGL\,J0658.6+0636. The observation campaign and its results are discussed in more details at this conference in \cite{201114A}.

\begin{table}
\small
\caption{List of IceCube singlet alerts followed up by at least one IACT. Energy and signalness estimates are not available for 200107A at the moment. IACT-named columns give the total exposure time (in hours) taken by each instrument.}
\label{tab_tracks}
\centering
\begin{tabular}{l|cccccc}
\hline
Name & Energy [TeV] & Signalness & FACT & H.E.S.S.\ & MAGIC & VERITAS\\
 \hline
IceCube-171106A & 230 & 0.75 & 19 h & --- & 4.5 h & 2.5 h \\
IceCube-181023A & 120 & 0.28 & 1 h & --- & --- & --- \\
IceCube-190503A & 100 & 0.36 & --- & --- & 0.5 h & --- \\
IceCube-190730A & 299 & 0.67 & ---  & --- & 3.1 h & --- \\
IceCube-190922B & 187 & 0.50 & 5.4 h & --- & 2.2 h & --- \\
IceCube-191001A & 217 & 0.59 & 2.0 & ---  & 2.3 h & 1.0 h \\
IceCube-200107A & --- & --- & ---  & --- & 2.7 h & 9.5 h \\
IceCube-200926A & 670 & 0.44 & --- & 1.3 h & 1.0 h& \\
IceCube-201007A & 683 & 0.88 & --- & 3.25 h & 0.5 h & ---  \\
IceCube-201114A & 214 & 0.56 & ---  & 14.5 h & 6 h & 7 h\\
IceCube-201222A & 186 & 0.53 & ---  & --- & --- & 1.0 h \\
 \end{tabular}
\end{table}

\begin{table}
\small
\caption{List of GFU alerts followed-up by IACTs in the years 2019-2020. The all-sky alert is shown separately given the much larger number of statistical trials involved in the search. IACT-named columns give the total exposure time (in hours) taken by each instrument.}
\label{tab_GFU}
\centering
\begin{tabular}{l|ccccc}
\hline
Source & Duration [days]& Pre-trial significance  & H.E.S.S & MAGIC & VERITAS\\
 \hline
MG1\,J181841+0903 
& Multiple alerts & $>$ 3.3 $\sigma$ & --- & 1.6 h & ---\\

1ES\,1312-423 & 0.26 & 3.4 $\sigma$ & 2.6 h & --- & --- \\
PMN\,J2016-09  & 0.01 & 3.6 $\sigma$  & --- & 0.9 h & --- \\
OP 313  
& Multiple alerts & $>$ 3.0 $\sigma$ & --- & 3.2 h & --- \\
OC\,457  & 0.30 & 3.3 $\sigma$ & --- & 2.5 h & --- \\
GB6\,J0316+0904 & 2.25 & 3.1 $\sigma$ & 6 h  & 1.9 h  & 1.0 h\\
\hline
All-sky alert (PMN\,J0325-1843)   & 3.67 & 5.1 $\sigma$ & ---  & 2.0 h & --- \\
 \end{tabular}
\end{table}

\small
\section*{Acknowledgments}
The following persons contributed to the FACT analyses: 
D.~Dorner (U. Würzburg), 
T.~Bretz (RWTH Aachen)
The members of the FACT Collaboration gratefully acknowledge financial support from the agencies and organizations listed here: 
\url{https://fact-project.org/collaboration/icrc2021_acknowledgements.html}

The following persons contributed to the IceCube analyses: T.~Kintscher, M.~Mallamaci, C.~Boscolo~Meneguolo.

The following persons contributed to the H.E.S.S. analyses: H.~Ashkar (IRFU), Y.~Becherini (LNU), M.~Cerruti (APC), C.~Hoischen (Uni. Potsdam), R.~Konno (Desy), F. Sch\"ussler (IRFU), M.~Senniappan (LNU). The members of the H.E.S.S. Collaboration gratefully acknowledge financial support from the agencies and organizations listed here:~\url{https://www.mpi-hd.mpg.de/hfm/HESS/pages/publications/auxiliary/HESS-Acknowledgements-2021.html}. FS acknowledges support for this work by the Programme National des Hautes Energies of CNRS/INSU with INP and IN2P3, co-funded by CEA and CNES.

The following persons contributed to the MAGIC analyses: M.~Artero (IFAE-BIST), A.~Berti (MPIfP),  H.~B\"okenkamp (U. Dortmund), A.~Fattorini (U. Dortmund), L.~Heckmann (MPIfP), S.~Mangano (CIEMAT), H.A.~Mondal (SINP), K.~Noda (ICRR), S.~Sakurai (ICRR), L.~di~Venere (INFN-Bari), I.~Viale (U. Padova),  S.~Yoo (U. Kyoto)
The members of the MAGIC Collaboration gratefully acknowledge financial support from the agencies and organizations listed here: ~\url{https://magic.mpp.mpg.de/acknowledgments_ICRC2021}

The members of the VERITAS Collaboration gratefully acknowledge financial support from the agencies and organizations listed here:~\url{https://veritas.sao.arizona.edu/}.



\bibliographystyle{ICRC}
\typeout{}
\bibliography{references}

\clearpage
\section*{Full Author List: FACT Collaboration, IceCube Collaboration, MAGIC Collaboration, H.E.S.S. Collaboration, VERITAS Collaboration}

%
%
%

\input{icecube}

\input{veritas}

\end{document}

%% file: icecube.tex
\clearpage
\section*{Full Author List: IceCube Collaboration}




\scriptsize
\noindent
R. Abbasi$^{17}$,
M. Ackermann$^{59}$,
J. Adams$^{18}$,
J. A. Aguilar$^{12}$,
M. Ahlers$^{22}$,
M. Ahrens$^{50}$,
C. Alispach$^{28}$,
A. A. Alves Jr.$^{31}$,
N. M. Amin$^{42}$,
R. An$^{14}$
% ...
\\

\noindent
% ...
$^{12}$ Universit{\'e} Libre de Bruxelles, Science Faculty CP230, B-1050 Brussels, Belgium \\
% ...
$^{14}$ Department of Physics and Laboratory for Particle Physics and Cosmology, Harvard University, Cambridge, MA 02138, USA \\
% ...
$^{17}$ Department of Physics, Loyola University Chicago, Chicago, IL 60660, USA \\
$^{18}$ Dept. of Physics and Astronomy, University of Canterbury, Private Bag 4800, Christchurch, New Zealand \\
% ...
$^{22}$ Niels Bohr Institute, University of Copenhagen, DK-2100 Copenhagen, Denmark \\
% ...
$^{28}$ D{\'e}partement de physique nucl{\'e}aire et corpusculaire, Universit{\'e} de Gen{\`e}ve, CH-1211 Gen{\`e}ve, Switzerland \\
% ...
$^{31}$ Karlsruhe Institute of Technology, Institute for Astroparticle Physics, D-76021 Karlsruhe, Germany  \\
% ...
$^{42}$ Bartol Research Institute and Dept. of Physics and Astronomy, University of Delaware, Newark, DE 19716, USA \\
% ...
$^{50}$ Oskar Klein Centre and Dept. of Physics, Stockholm University, SE-10691 Stockholm, Sweden \\
% ...
$^{59}$ DESY, D-15738 Zeuthen, Germany \\
% ...

\subsection*{Acknowledgements}

\noindent
USA {\textendash} U.S. National Science Foundation-Office of Polar Programs,
U.S. National Science Foundation-Physics Division,
U.S. National Science Foundation-EPSCoR,
Wisconsin Alumni Research Foundation,
Center for High Throughput Computing (CHTC) at the University of Wisconsin{\textendash}Madison,
Open Science Grid (OSG),
Extreme Science and Engineering Discovery Environment (XSEDE),
Frontera computing project at the Texas Advanced Computing Center,
U.S. Department of Energy-National Energy Research Scientific Computing Center,
Particle astrophysics research computing center at the University of Maryland,
Institute for Cyber-Enabled Research at Michigan State University,
and Astroparticle physics computational facility at Marquette University;
Belgium {\textendash} Funds for Scientific Research (FRS-FNRS and FWO),
FWO Odysseus and Big Science programmes,
and Belgian Federal Science Policy Office (Belspo);
Germany {\textendash} Bundesministerium f{\"u}r Bildung und Forschung (BMBF),
Deutsche Forschungsgemeinschaft (DFG),
Helmholtz Alliance for Astroparticle Physics (HAP),
Initiative and Networking Fund of the Helmholtz Association,
Deutsches Elektronen Synchrotron (DESY),
and High Performance Computing cluster of the RWTH Aachen;
Sweden {\textendash} Swedish Research Council,
Swedish Polar Research Secretariat,
Swedish National Infrastructure for Computing (SNIC),
and Knut and Alice Wallenberg Foundation;
Australia {\textendash} Australian Research Council;
Canada {\textendash} Natural Sciences and Engineering Research Council of Canada,
Calcul Qu{\'e}bec, Compute Ontario, Canada Foundation for Innovation, WestGrid, and Compute Canada;
Denmark {\textendash} Villum Fonden and Carlsberg Foundation;
New Zealand {\textendash} Marsden Fund;
Japan {\textendash} Japan Society for Promotion of Science (JSPS)
and Institute for Global Prominent Research (IGPR) of Chiba University;
Korea {\textendash} National Research Foundation of Korea (NRF);
Switzerland {\textendash} Swiss National Science Foundation (SNSF);
United Kingdom {\textendash} Department of Physics, University of Oxford.

%% file: veritas.tex
\clearpage
\section*{Full Authors List: VERITAS Collaboration}

\scriptsize
\noindent
C.~B.~Adams$^{1}$,
A.~Archer$^{2}$,
W.~Benbow$^{3}$,
A.~Brill$^{1}$,
J.~H.~Buckley$^{4}$,
M.~Capasso$^{5}$,
J.~L.~Christiansen$^{6}$,
A.~J.~Chromey$^{7}$, 
M.~Errando$^{4}$,
A.~Falcone$^{8}$,
K.~A.~Farrell$^{9}$,
Q.~Feng$^{5}$,
G.~M.~Foote$^{10}$,
L.~Fortson$^{11}$,
A.~Furniss$^{12}$,
A.~Gent$^{13}$,
G.~H.~Gillanders$^{14}$,
C.~Giuri$^{15}$,
O.~Gueta$^{15}$,
D.~Hanna$^{16}$,
O.~Hervet$^{17}$,
J.~Holder$^{10}$,
B.~Hona$^{18}$,
T.~B.~Humensky$^{1}$,
W.~Jin$^{19}$,
P.~Kaaret$^{20}$,
M.~Kertzman$^{2}$,
T.~K.~Kleiner$^{15}$,
S.~Kumar$^{16}$,
M.~J.~Lang$^{14}$,
M.~Lundy$^{16}$,
G.~Maier$^{15}$,
C.~E~McGrath$^{9}$,
P.~Moriarty$^{14}$,
R.~Mukherjee$^{5}$,
D.~Nieto$^{21}$,
M.~Nievas-Rosillo$^{15}$,
S.~O'Brien$^{16}$,
R.~A.~Ong$^{22}$,
A.~N.~Otte$^{13}$,
S.~R. Patel$^{15}$,
S.~Patel$^{20}$,
K.~Pfrang$^{15}$,
M.~Pohl$^{23,15}$,
R.~R.~Prado$^{15}$,
E.~Pueschel$^{15}$,
J.~Quinn$^{9}$,
K.~Ragan$^{16}$,
P.~T.~Reynolds$^{24}$,
D.~Ribeiro$^{1}$,
E.~Roache$^{3}$,
J.~L.~Ryan$^{22}$,
I.~Sadeh$^{15}$,
M.~Santander$^{19}$,
G.~H.~Sembroski$^{25}$,
R.~Shang$^{22}$,
D.~Tak$^{15}$,
V.~V.~Vassiliev$^{22}$,
A.~Weinstein$^{7}$,
D.~A.~Williams$^{17}$,
and 
T.~J.~Williamson$^{10}$\\

\noindent

$^1${Physics Department, Columbia University, New York, NY 10027, USA}
$^{2}${Department of Physics and Astronomy, DePauw University, Greencastle, IN 46135-0037, USA}
$^3${Center for Astrophysics $|$ Harvard \& Smithsonian, Cambridge, MA 02138, USA}
$^4${Department of Physics, Washington University, St. Louis, MO 63130, USA}
$^5${Department of Physics and Astronomy, Barnard College, Columbia University, NY 10027, USA}
$^6${Physics Department, California Polytechnic State University, San Luis Obispo, CA 94307, USA} 
$^7${Department of Physics and Astronomy, Iowa State University, Ames, IA 50011, USA}
$^8${Department of Astronomy and Astrophysics, 525 Davey Lab, Pennsylvania State University, University Park, PA 16802, USA}
$^9${School of Physics, University College Dublin, Belfield, Dublin 4, Ireland}
$^{10}${Department of Physics and Astronomy and the Bartol Research Institute, University of Delaware, Newark, DE 19716, USA}
$^{11}${School of Physics and Astronomy, University of Minnesota, Minneapolis, MN 55455, USA}
$^{12}${Department of Physics, California State University - East Bay, Hayward, CA 94542, USA}
$^{13}${School of Physics and Center for Relativistic Astrophysics, Georgia Institute of Technology, 837 State Street NW, Atlanta, GA 30332-0430}
$^{14}${School of Physics, National University of Ireland Galway, University Road, Galway, Ireland}
$^{15}${DESY, Platanenallee 6, 15738 Zeuthen, Germany}
$^{16}${Physics Department, McGill University, Montreal, QC H3A 2T8, Canada}
$^{17}${Santa Cruz Institute for Particle Physics and Department of Physics, University of California, Santa Cruz, CA 95064, USA}
$^{18}${Department of Physics and Astronomy, University of Utah, Salt Lake City, UT 84112, USA}
$^{19}${Department of Physics and Astronomy, University of Alabama, Tuscaloosa, AL 35487, USA}
$^{20}${Department of Physics and Astronomy, University of Iowa, Van Allen Hall, Iowa City, IA 52242, USA}
$^{21}${Institute of Particle and Cosmos Physics, Universidad Complutense de Madrid, 28040 Madrid, Spain}
$^{22}${Department of Physics and Astronomy, University of California, Los Angeles, CA 90095, USA}
$^{23}${Institute of Physics and Astronomy, University of Potsdam, 14476 Potsdam-Golm, Germany}
$^{24}${Department of Physical Sciences, Munster Technological University, Bishopstown, Cork, T12 P928, Ireland}
$^{25}${Department of Physics and Astronomy, Purdue University, West Lafayette, IN 47907, USA}